\documentclass{aa}
\usepackage{txfonts,natbib,graphicx,units}
\bibpunct{(}{)}{;}{a}{}{,}

\begin{document}
\title{The evolutionary status of the bright\\high-latitude supergiant
HD\,190390\thanks{based on observations collected at the
European Southern Observatory, La Silla, Chile (programme 61.E-0426), and at
the Observatorio del Roque de los Muchachos, La Palma, Spain}}
\author{Maarten Reyniers\inst{1,}\thanks{Postdoctoral fellow of the Fund for
Scientific Research, Flanders} \and Jan Cuypers\inst{2}}
\institute{Instituut voor Sterrenkunde, Departement Natuurkunde en Sterrenkunde,
K.U.Leuven, Celestijnenlaan 200B, 3001 Leuven, Belgium
\and
Koninklijke Sterrenwacht van Belgi\"e, Ringlaan 3, 1180 Brussel, Belgium}
\offprints{Maarten Reyniers, \email{Maarten.Reyniers@ster.kuleuven.ac.be}}
\date{Received 30 September 2004 / Accepted 31 October 2004}
\authorrunning{M. Reyniers \& J. Cuypers}
\titlerunning{The evolutionary status of HD\,190390}

\abstract{Despite its mean apparent magnitude of $m_{\rm V}$\,=\,6.39, the
evolutionary status of \object{HD\,190390} (\object{HR\,7671}), a luminous
F-type supergiant at high galactic latitude, is still not very clear, but in
most papers a post-AGB classification is assumed. New observational material
has been obtained with four different instruments and is presented here. An
extensive abundance analysis based on high resolution, high signal-to-noise
NTT+EMMI spectra confirms the metal deficiency of this object
([Fe/H]\,=\,$-$1.6), together with a high lithium content
($\log \epsilon$(Li)\,=\,1.9). A variability analysis based on Geneva
photometry over seven years reveals beating with a period of $\sim$3000 days.
It is, however, not clear whether this beating is caused by a stable triplet,
or it is the consequence of small changes in the main frequency. More recent
data obtained with the HIPPARCOS satellite and the Mercator telescope not only
confirm the main period, but also support the presence of a second periodicity
of 11 days, which was also found in the Geneva photometry. A conclusive
evolutionary status of this object is not given, but alternative to the UU~Her
(i.e. post-AGB) status, a W~Vir classification is discussed.
}
\maketitle

\section{Introduction}\label{sect:ntrdctn}
Since their discovery by William \citet{bidelman51}, the evolutionary stage
of \object{HR\,6144} (b\,=\,$+$25$^{\circ}$), \object{89\,Her}
($+$22$^{\circ}$) and \object{HD\,161796} ($+$30$^{\circ}$), three
high-luminosity stars at moderately high galactic latitude, has been a matter
of debate. At least three different possibilities were suggested: (1) young,
massive objects that recently escaped from a star forming region, (2) low-mass,
evolved objects or (3) a rare product in a binary star scenario. After the high
resolution, high signal-to-noise spectroscopic study of \citet*{luck90}, there
was general consensus that these objects are low-mass objects in an evolved,
post-AGB evolutionary stage. Although {\em no} s-process enhancement was
observed in these stars, there were enough indications for a classification as
a post-AGB star (slightly metal-poor, overabundant carbon and oxygen and a
strong infrared dust excess indicating high mass loss in a previous
evolutionary (AGB) stage).

\begin{table}\caption{Basic parameters of \object{HD\,190390} (SIMBAD database).}\label{tab:smbdgegevens}
\begin{center}
\begin{tabular}{rrr}
\hline
\multicolumn{3}{c}{\rule[-0mm]{0mm}{4mm}\bf\large \object{HD\,190390} (\object{HR\,7671})}\\[.4mm]
\hline
Coordinates   & $\alpha_{2000}$ & 20 05 05.41 \\
              & $\delta_{2000}$ &$-$11 35 57.9\phantom{0} \\
\hline Galactic    & $l$ &30.60  \\ 
       coordinates & $b$ &$-$21.53 \\ 
\hline
Mean magnitude   &   B &6.88 \\
                 &   V &6.39 \\
\hline
Spectral type$^*$ &  & F1III \\
\hline
Parallax      &$\pi$\,(mas) & 2.56\,$\pm$\,0.97\\
\hline
IRAS fluxes   &$f_{12}$  & 0.65\\
       (Jy)   &$f_{25}$  & $<$\,0.32\\
              &$f_{60}$  & $<$\,0.40\\
              &$f_{100}$ & $<$\,1.08\\
\hline
\end{tabular}
\end{center}
{\small $^*$\,very differing spectral types have been reported in the
literature: gF4, F1III, F6II, F6Ib, F3Ib, F2p\,(shell) \citep[cit.][]{fernie86}}
\end{table}

There was also a fourth star in the \citeauthor{luck90} sample,
\object{HD\,190390} (\object{HR\,7671}, Table~\ref{tab:smbdgegevens}). The
inclusion of this object in their sample was motivated not only by its similar
galactic latitude and spectral type, but also by the variability pointed out by
\citet{waelkens85} and \citet{fernie86}. This variablity made
\object{HD\,190390} a candidate member of the UU~Her type stars, a
heterogeneous class of luminous variables at high galactic latitudes defined by
\citet{sasselov84}, which also includes \object{89\,Her} and
\object{HD\,161796}. For \object{HD\,190390}, however, \citet{luck90} found
a clearly different chemical signature than for the other three stars, with a
C and O deficiency and a high Li abundance. The authors proposed the object to
be a descendant of a Li-rich S-type star, with the Li attributed to a ``hot
bottom burning'' event.

In this paper, new observational material is presented to clarify the
status of this intriguing object. Both the chemical composition and the
variability are reanalysed, based on high quality data coming from different
instruments and telescopes. In Sect.~\ref{sect:chmclcomposition} we report on
the reduction and analysis of our high resolution, high signal-to-noise
NTT+EMMI spectra and discuss the chemical photospheric composition of
\object{HD\,190390}. In Sect.~\ref{sect:vrbltanalysis}, a detailed variability
analysis is presented, based on both photometric data and radial velocities.
Some results are presented originating from the new Flemish Mercator telescope
located at La Palma, Spain. Our results are discussed in
Sect.~\ref{sect:dscssn}. We give our conclusions in Sect.~\ref{sect:cnclsns}.

\section{Chemical composition}\label{sect:chmclcomposition}
\subsection{Previous studies}
\citet*{luck90} were the first to carry out a detailed abundance analysis
of \object{HD\,190390}, based on spectra with medium to high resolution
($R=\lambda/\delta\lambda$\,$\simeq$\,15000--47000) and a signal-to-noise ratio
between 75 and 100. They confirmed the metal deficiency discovered by
\citet{mcdonald76} and quantified it: [Fe/H]\,=\,$-$1.1. Further, they found a
slight C and O deficiency and also claimed a slight s-process enhancement by a
factor of 4 above solar. Lithium was also detected with an abundance of
log\,$\epsilon$(Li)\,=\,2.4. The main shortcomings of this analysis were the
quite low signal-to-noise ratio obtained, and the inhomogeneous collection of
spectra with different resolution, leading to quite large line-to-line scatters
of the reported abundances (up to 0.92\,dex). A more recent abundance analysis
was presented by \citet*{giridhar97}. This analysis was based on moderately
high resolution spectra ($R\simeq30000$) with a rather narrow wavelength
coverage (5260--5600\,\AA), and it is therefore rather limited.

In the analysis we present here, we avoided the shortcomings of the previous
studies: a large spectral domain covered with the same resolution and a high
signal-to-noise was used. Such spectra were taken with ESO-NTT telescope and
the EMMI spectrograph.

\subsection{Observations and reduction}\label{subsect:obsrvtns}
\begin{figure*}
\resizebox*{\hsize}{!}{\rotatebox{-90}{\includegraphics{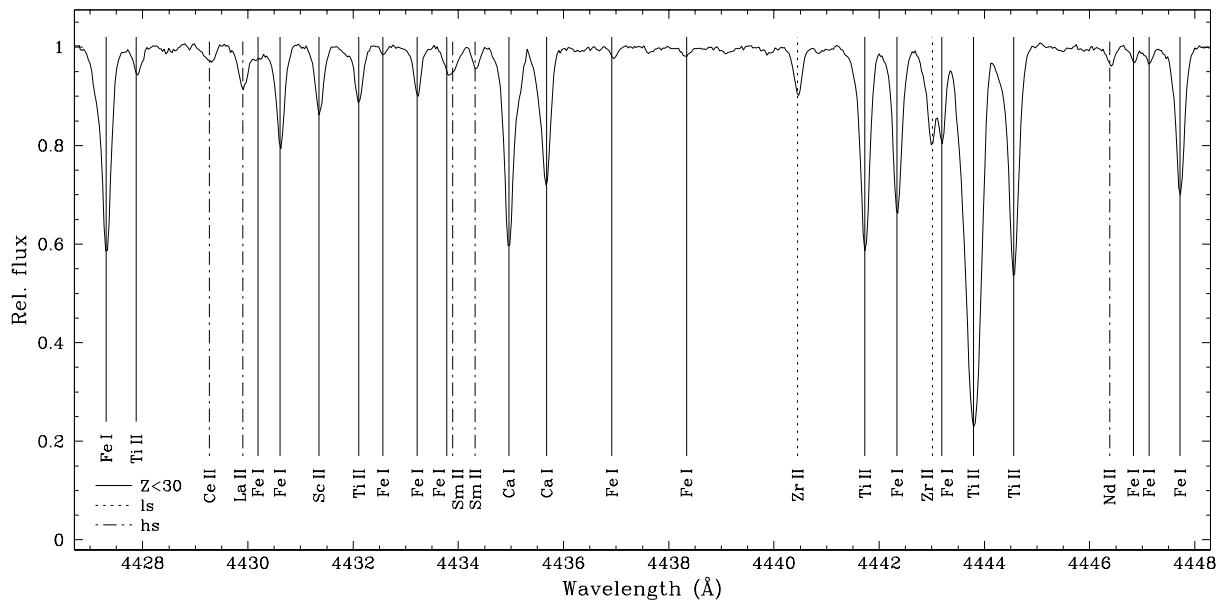}}}
\caption{The spectrum of \object{HD\,190390} between 4425 and 4450\,\AA. A
complete line identification of this spectral interval has been attempted by
the use of the VALD database \citep{kupka99}. Lines of light s-process elements
(Sr peak) are shown with a dotted line; lines of heavy s-process elements (Ba
peak) with a dash-dotted line; lines of other elements (which are mainly
$\alpha$ and iron peak elements) with a full line. The signal-to-noise ratio of
our spectra is around 500.}\label{fig:figr2004}
\end{figure*}

In the framework of our ongoing program to study the photospheric chemical
composition of stars in their last stages of evolution
\citep[e.g.][]{vanwinckel03, reyniers04}, high resolution, high
signal-to-noise spectra were taken with the EMMI Spectrograph mounted on the
3.58\,m New Technology Telescope (NTT) located in La Silla, Chile. The
observations of \object{HD\,190390} were made by Dr. H. Van Winckel during two
nights in 1998 (see Table~\ref{tab:obsdtls}). The reduction of the data was
performed in the specific echelle package {\sc echelle} of the {\sc midas} data
analysis system. For a more detailed description of the different steps of the
reduction procedure, see \citet{reyniers02a}. A sample spectrum can be found in
Fig.~\ref{fig:figr2004}.

\begin{table}\caption{Log of the high resolution, high signal-to-noise NTT+EMMI
spectra of \object{HD\,190390}. The resolving power $\lambda/\delta\lambda$ was
$\sim$60000 for both settings.}\label{tab:obsdtls}
\begin{center}
\begin{tabular}{rrrr}
\hline
\multicolumn{1}{c}{Date and UT} & \multicolumn{1}{c}{EMMI}    & \multicolumn{1}{c}{Integration} &  \multicolumn{1}{c}{Sp. Range} \\
                                & \multicolumn{1}{c}{setting} & \multicolumn{1}{c}{Time (s)}    &  \multicolumn{1}{c}{(nm)}      \\
\hline
29/09/1998 02:55 & Ech\#14 Grsm\#5 &  900 & 398-662 \\
30/09/1998 04:18 & Ech\#14 Grsm\#6 & 1200 & 597-832 \\
\hline
\end{tabular}
\end{center}
\end{table}

\subsection{Abundance analysis}\label{sect:abndc}
A detailed abundance analysis of \object{HD\,190390} was performed with an
extended version of the line list described in \citet{vanwinckel00}.

\subsubsection{Atmospheric parameters and equivalent widths}
\begin{figure}
\resizebox{\hsize}{!}{\includegraphics{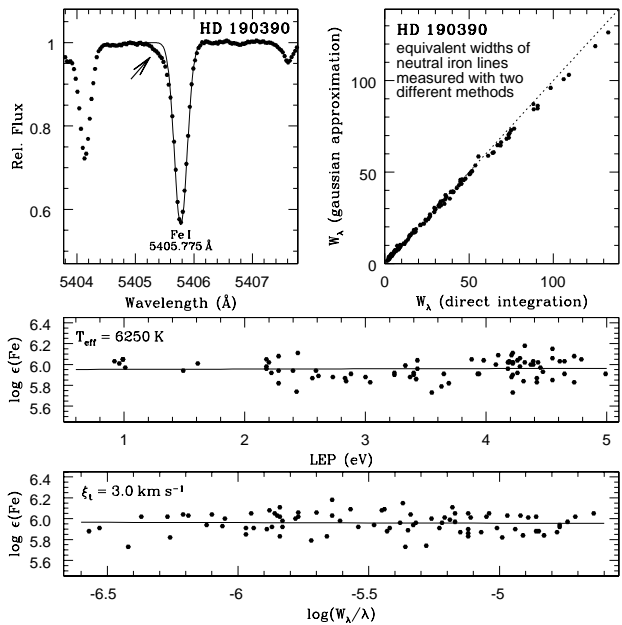}}
\caption{Lines of neutral iron in the spectrum of \object{HD\,190390}. The small
asymmetry in the blue wings of the stronger lines makes a Gaussian
approximation for these lines not acceptable. This is illustrated in the
{\em upper panels} of the figure. Taking the Gaussian equivalent widths would
lead to a wrong $\xi_t$. The {\em two lower panels} are E.P.\,-\,abundance
and $\log(W_{\lambda}/\lambda$)\,-\,abundance diagrams for the final model of
\object{HD\,190390}.}\label{fig:abndcplots}
\end{figure}

Since the atmospheric parameters are usually based on a fine analysis of the
iron lines, the \citeauthor{vanwinckel00} list only contains the critical
compilation of Lambert \citep{lambert96}, completed with values of
\citet{blackwell80} for the singly ionised iron lines. Two different methods
were used to measure the equivalent widths: a Gaussian approximation and direct
integration. Due to a small asymmetry in the blue wing of the stronger lines,
a Gaussian approximation leads to an underestimation of the real equivalent
width. This is illustrated in the top panels of Fig.~\ref{fig:abndcplots}. On
the other hand, a Gaussian approximation allows a better continuum placement
(especially important for weaker lines) and also allows for blends in the wing
of a line. We have chosen a strictly homogeneous way of measuring the
equivalent widths of the iron lines: for lines with
$W_{\lambda}$\,$<$\,10.5\,m\AA, we applied a Gaussian approximation, for the
other lines we used direct integration, except if the line was blended in the
wings. 84 Fe\,{\sc i} lines and 24 Fe\,{\sc ii} lines were measured in such way.

The effective temperature T$_{\rm eff}$ of the Kurucz model atmosphere
\citep{kurucz93} was then obtained by forcing the Fe abundance to be
independent of the excitation potential of the lines, the gravity $\log g$ by
forcing ionisation equilibrium, the microturbulent velocity $\xi_t$ by forcing
the Fe abundance to be independent of (reduced) equivalent widths. Abundance
calculations were made with the LTE line analysis program MOOG (version April
2002). This iterative process finally yielded a model with
(T$_{\rm eff}$,\,$\log g$,\,$\xi_t$,\,[M/H]) = 
(6250\,K,\,1.0\,(cgs),\,3.0\,km\,s$^{-1}$,\,$-$1.5). E.P.\,-\,abundance and
$\log(W_{\lambda}/\lambda$)\,-\,abundance diagrams for this model can be found
in Fig.~\ref{fig:abndcplots}. The model parameters slightly differ from the
model parameters found by \citet{luck90} and \citet{giridhar97}. The former
authors found (6600\,K,\,1.5\,(cgs),\,2.3\,km\,s$^{-1}$), the latter 
(6500\,K,\,1.25\,(cgs),\,3.2\,km\,s$^{-1}$). However, due to the superior
quality of our spectra and the severe selection of our Fe line list, we feel
confident about our slightly differing parameters. Note that the difference
in atmospheric parameters can also be a consequence of the variability of
the object.

\begin{figure}
\resizebox{\hsize}{!}{\includegraphics{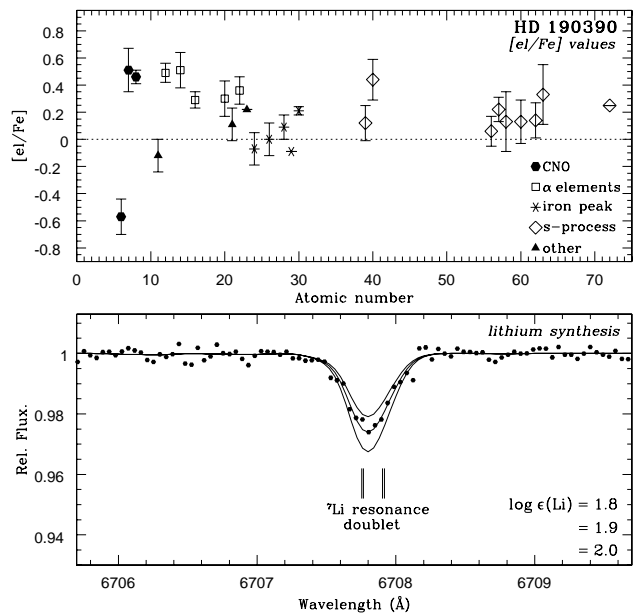}}
\caption{[el/Fe] values ({\em upper panel}) and lithium doublet synthesis
({\em lower panel}) for \object{HD\,190390}. The error on the [el/Fe] values
is the line-to-line scatter found in Table~\ref{tab:abndctabel}.}\label{fig:cnoelfedefic}
\end{figure}

The method of the equivalent width measurement (either Gaussian approximation
or direct integration) for the other lines was chosen for each line
individually, without a strict criterion. The \citeauthor{vanwinckel00} list
was extended for the s-process elements in the following way. We extracted all
s-process elements from the VALD database \citep{kupka99}. These lists were
then sorted based on the calculated equivalent widths of the lines using the
model obtained above and an ad-hoc abundance value. Each line of these lists
was then checked in this order on the spectrum of \object{HD\,190390}, until
the lines were too small to detect. 136 lines of s-process elements were used
in the analysis. The oscillator strengths of La and Eu from VALD were replaced
by the more recent values published by \citet{lawler01a} and \citet{lawler01b}
respectively. The total number of spectral lines used in this analysis is 462.
The line list can be obtained from the authors upon request.

\subsubsection{Abundance results}\label{subsubs:abundancerslts}
The resulting abundances are given in Table~\ref{tab:abndctabel}. For each
ion we list the number of lines N, the mean equivalent width
$\overline{W_{\lambda}}$, the absolute abundance $\log\epsilon$ (i.e. relative
to H: $\log\epsilon = \log X/{\rm H} + 12$), the internal scatter $\sigma$, if
more than one line is used, the solar abundances and the abundance ratio
relative to iron [el/Fe]. The latter values are also graphically presented in
the top panel of Fig.~\ref{fig:cnoelfedefic}. For the solar iron abundance we
used the meteoritic iron abundance of 7.51; the references for the solar CNO
are C: \citealt{biemont93}, N: \citealt{hibbert91} and O: \citealt{biemont91};
for Mg and Si the latest Holweger values \citep{holweger01} were used; for La
and Eu we took the recent values derived by \citet{lawler01a} and
\citet{lawler01b} respectively; other solar abundances were taken from the
review by \citet{grevesse98}. Despite the fact that there are more recent
values for some of the solar abundances (especially for the solar CNO), we take
these references to ensure as much as possible that the $gf$ values that we
have used in the present paper are consistent with the adopted solar
abundances. Note the excellent agreement in abundance for elements with two
ionisation stages, Mg, Si, Ti, Cr and Ni, indicating a good model parameter
choice. Note also the small line-to-line scatter which rarely exceeds
0.15\,dex, including the scatter for the s-process ions.

\begin{table}\caption{Abundance results for \object{HD\,190390}. The references
for the solar abundances are given in the text (Sect.~\ref{subsubs:abundancerslts}).}\label{tab:abndctabel}
\begin{center}
\begin{tabular}{lrrrrrr}
\hline
\multicolumn{7}{c}{\rule[-0mm]{0mm}{5mm}{\Large\bf \object{HD\,190390}}}\\
\multicolumn{7}{c}{
$
\begin{array}{r@{\,=\,}l}
{\rm T}_{\rm eff} & 6250\,{\rm K}  \\
\log g  & 1.0\ {\rm(cgs)} \\
\xi_{\rm t} & 3.0\ {\rm km\,s}^{-1} \\
\end{array}
$
}\\
\hline
ion & N &{\rule[0mm]{0mm}{4mm}$\overline{W_{\lambda}}$}&$\log\epsilon$&$\sigma$& sun &[el/Fe]\\
\hline
Li\,{\sc i}  &  1 & ss &  1.9\phantom{0}&0.2\phantom{0}& &   \\
\hline
C\,{\sc i}   &  9 &  4 &  6.44 & 0.13 & 8.57 &$-$0.57\\
N\,{\sc i}   &  3 &  5 &  6.94 & 0.16 & 7.99 &  0.51\\
O\,{\sc i}   &  3 &  5 &  7.76 & 0.05 & 8.86 &  0.46\\
\hline
Na\,{\sc i}  &  2 & 11 &  4.65 & 0.12 & 6.33 &$-$0.12\\
Mg\,{\sc i}  &  3 & 91 &  6.47 & 0.09 & 7.54 &  0.49\\
Mg\,{\sc ii} &  1 &  8 &  6.46 &      & 7.54 &  0.48\\
Si\,{\sc i}  & 17 & 11 &  6.50 & 0.12 & 7.54 &  0.52\\
Si\,{\sc ii} &  5 & 46 &  6.47 & 0.17 & 7.54 &  0.49\\
S\,{\sc i}   &  5 &  9 &  6.06 & 0.06 & 7.33 &  0.29\\
Ca\,{\sc i}  & 23 & 36 &  5.10 & 0.13 & 6.36 &  0.30\\
Sc\,{\sc ii} & 13 & 54 &  1.72 & 0.12 & 3.17 &  0.11\\
Ti\,{\sc i}  &  2 & 27 &  3.80 & 0.01 & 5.02 &  0.34\\
Ti\,{\sc ii} & 42 & 76 &  3.82 & 0.10 & 5.02 &  0.36\\
V\,{\sc ii}  &  1 & 96 &  2.66 &      & 4.00 &  0.22\\
Cr\,{\sc i}  & 16 & 18 &  4.07 & 0.13 & 5.67 &$-$0.04\\
Cr\,{\sc ii} & 23 & 28 &  4.01 & 0.11 & 5.67 &$-$0.10\\
Mn\,{\sc i}  &  9 &  9 &  3.52 & 0.10 & 5.39 &$-$0.31\\
Fe\,{\sc i}  & 84 & 30 &  5.96 & 0.10 & 7.51 &  0.01\\
Fe\,{\sc ii} & 24 & 41 &  5.95 & 0.12 & 7.51 &  0.00\\
Ni\,{\sc i}  & 35 & 10 &  4.78 & 0.10 & 6.25 &  0.09\\
Ni\,{\sc ii} &  2 &  7 &  4.79 & 0.01 & 6.25 &  0.10\\
Cu\,{\sc i}  &  1 &  3 &  2.56 &      & 4.21 &$-$0.09\\
Zn\,{\sc i}  &  3 & 19 &  3.25 & 0.03 & 4.60 &  0.21\\
\hline
Y\,{\sc ii}  & 22 & 34 &  0.80 & 0.13 & 2.24 &  0.12\\
Zr\,{\sc ii} & 31 & 35 &  1.48 & 0.15 & 2.60 &  0.44\\
Ba\,{\sc ii} &  3 & 30 &  0.63 & 0.11 & 2.13 &  0.06\\
La\,{\sc ii} & 16 & 19 &$-$0.21& 0.09 & 1.13 &  0.22\\ 
Ce\,{\sc ii} & 32 & 11 &  0.15 & 0.22 & 1.58 &  0.13\\
Nd\,{\sc ii} & 14 & 10 &  0.07 & 0.16 & 1.50 &  0.13\\
Sm\,{\sc ii} & 15 &  6 &$-$0.41& 0.13 & 1.01 &  0.14\\
Eu\,{\sc ii} &  2 & ss &$-$0.71& 0.22 & 0.52 &  0.33\\ 
Hf\,{\sc ii} &  1 &  4 &$-$0.43&      & 0.88 &  0.25\\
\hline 
\multicolumn{7}{l}{[Fe/H]\,=\,$-$1.56}\\
\multicolumn{7}{l}{C/O\,=\,0.05}\\
\multicolumn{7}{l}{[$\alpha$/Fe]\,=\,$+$0.4 ($\alpha$: Mg, Si, S, Ca, Ti)}\\
\hline
\end{tabular}
\end{center}
\end{table}


\subsubsection{Lithium synthesis}
The lithium line in \object{HD\,190390} with equivalent width
$W_{\lambda}$\,=\,9.5\,m\AA\ was first reported by \citet{luck90}. To account
for the doublet structure of the line, spectrum synthesis is necessary
in deriving the Li abundance. The program MOOG was also used for this purpose.
The line list is identical to the line list that was used in
\citet{reyniers02b}, but due to the combined effect of the quite high
temperature and the metal deficiency of \object{HD\,190390}, no other lines are
detectable in the 6707\,\AA\ region.

For a spectral synthesis, a broadening factor is needed, in addition to the
model atmosphere parameters. This broadening factor is a combined effect of
the instrumental, rotational and macroturbulent broadening. An instrumental
broadening of 5\,km\,s$^{-1}$ was adopted (the spectral resolution of the
EMMI spectrograph). The rotational broadening is thought not to be very high
due to the supergiant character of \object{HD\,190390}. The macroturbulent
broadening $\xi_m$ was determined in the following way. We selected 27
unblended iron lines (neutral and ionised) with $W_{\lambda}$\,$<$\,25\,m\AA.
For each of these lines, we made a synthesis using the abundance calculated
via their $W_{\lambda}$. In such way, the macroturbulent broadening is the
only free parameter in this synthesis. We found a quite large spread in
$\xi_m$ for these lines: $\xi_m$\,=\,8.0\,$\pm$\,1.2\,km\,s$^{-1}$. This is
also exactly the value found in the \citeauthor{luck90} analysis. Hence for
the lithium synthesis $\xi_m$\,=\,8.0\,km\,s$^{-1}$ was used. This yielded
log\,$\epsilon$(Li)\,=\,1.9.

An error analysis on the Li abundance was made by varying all parameters
involved in the synthesis. As expected, a temperature change has far the largest
impact on the abundance: a temperature variation of $\pm$250\,K yields a change
in the Li abundance of $\sim$0.2 dex. Other variations are listed in
Table~\ref{tab:lthmvaria}.

\subsubsection{Europium synthesis}
The effect of hyperfine splitting ({\em hfs}) on the europium abundance was
investigated by the synthesis of the two Eu\,{\sc ii} lines of our analysis.
$\log gf$ data and hyperfine splitting constants were taken from
\citet{lawler01b}. The influence of hfs was determined by integrating the
profile of the synthesized line, with and without hfs decomposition. This
calculated equivalent width is then compared with the measured equivalent
width. In this way, the exact profile of the line is eliminated, and no
broadening factors have to be applied. The effect on abundance of hfs for the
resonance line at 4129.725\,\AA\ (W$_{\lambda}$\,=\,43.1\,m\AA) is significant.
A hfs treatment of this line yields $\log\epsilon({\rm Eu})$\,=\,$-$0.87\,dex, 
while without hfs it is $-$0.79\,dex. For the weak line at 6645.064\,\AA\ 
(W$_{\lambda}$\,=\,5.9\,m\AA), the difference between an hfs and a non-hfs
treatment is not significant.

\subsection{Abundances: summary}
Referring to Fig.~\ref{fig:cnoelfedefic}, we will now discuss the derived
abundances of \object{HD\,190390}.\\
\noindent {\bf Metallicity}
With an iron abundance of log\,$\epsilon({\rm Fe})$\,=\,5.95,
\object{HD\,190390} is clearly metal deficient by almost a factor of 40. The
other iron peak elements follow this deficiency within 0.1\,dex, except Mn and
Zn. The slight manganese deficiency ([Mn/Fe]\,=\,$-$0.3) is expected given the
galactic chemical evolution \citep[e.g.][]{nissen00, prochaska00}. The zinc
abundance ([Zn/Fe]\,=\,$+$0.2) is somewhat higher than expected, but it is
based on only three lines.\\
{\bf CNO}
Although the abundance of the CNO elements is solely based on very small
lines, a clear pattern for these three elements is observed: [C/Fe]\,=\,$-$0.6
and [N/Fe]\,=\,[O/Fe]\,=\,$+0.5$. The deficient carbon and the overabundant
oxygen lead to a very low C/O number ratio: C/O\,=\,0.05. Nitrogen seems to
be enhanced at the same level as oxygen.\\
{\bf \boldmath$\alpha$-elements}
The mean of the [el/Fe] values of the $\alpha$-elements Mg, Si, S, Ca and Ti
yields [$\alpha$/Fe]\,= $+$0.4. This enhancement is what is expected for stars
in this metallicity range and hence the $\alpha$-elements are {\em not}
intrinsically enhanced. The abundances of the individual $\alpha$-elements
all agree within 0.1\,dex with this $+$0.4 number.\\
{\bf S-process elements}
A very complete abundance pattern could be obtained for the s-process elements,
with 53 lines of the light ({\em ls}) and 83 lines of the heavy ({\em hs})
s-process elements. The s-process elements are all slightly enhanced, but
the abundance pattern reveals that the enhancement is probably {\em not}
intrinsic. All elements are compatible with [el/Fe]\,=\,0 considering the
line-to-line scatters, except Zr, La, Eu and Hf. For Zr, a large spread
is seen in the [Zr/Fe] of unevolved disk and halo stars from $-$0.1 to
$+$0.6\,dex \citep[e.g.][]{travaglio04}. Moreover, the same relatively high Zr
abundance is also seen in our analysis of the non-enriched RV\,Tauri star
DS\,Aqr \citep{deroo05}. Eu is an element with a clear r-process origin, and
an enhancement of [Eu/Fe]\,=\,$+$0.4 is certainly expected in this metallicity
range \citep[e.g.][]{travaglio99}. The Hf abundance is based on only one very
small line and should therefore be interpreted with care. Moreover, the
galactic chemical evolution of this very heavy element is still unknown.
Finally, La seems to be a little bit more enhanced than the other s-process
elements, but still compatible with {\em no} intrinsic enhancement regarding
the scatter (although not as pronounced as for Zr) for unevolved halo stars
\citep[e.g.][]{travaglio99}. This result is contrary to the s-process
enrichment claimed by \citet{luck90}. They found a mild enhancement by a
factor of 4 compared to solar. A detailed inspection of their results,
however, reveals large line-to-line scatters for all elements ($>$0.25,
except for Ba) and a smaller number of lines per element. Also, the abundances
of the s-process elements as determined by \citeauthor{luck90} do not
display a consistent pattern in the sense that some elements are much more
enhanced than others. [Sm/Fe]=$+$2.1 together with [Ce/Fe]=$+$0.2 is
theoretically very difficult to reconcile with an s-process signature. We
conclude that the s-process abundances presented here are much more reliable
than the ones from the \citeauthor{luck90} paper.\\
{\bf Lithium}
The lithium abundance found in this analysis (log\,$\epsilon$(Li)\,=\,1.9)
is 0.5\,dex lower than that found in the \citeauthor{luck90} analysis. The
reason is the higher effective temperature in the latter analysis. Indeed,
taking a model with a temperature of 6500\,K in the Li synthesis, our derived
Li abundance rises by almost 0.2\,dex.\\
{\bf Other elements}
Sodium is slightly underabundant, but not at a significant level
\citep[e.g.][]{carretta00}. The slight scandium enhancement
([Sc/Fe]\,=\,$+$0.1) is not unusual in this metallicity range
\citep[e.g.][]{nissen00, prochaska00}. Finally, the vanadium (V) abundance
is quite high, but it is based on only one line.

\begin{table}\caption{Error analysis of the Li synthesis. A temperature change
has the largest effect on the derived Li abundance.}\label{tab:lthmvaria}
\begin{center}
\begin{tabular}{rrlr}
\hline
parameter             & symbol              & variation & variation \\
                      &                     & &$\log\epsilon({\rm Li})$ \\
\hline
temperature &T$_{\rm eff}$& $\pm$250\,K     & $\pm$0.20\\
gravity     & $\log(g)$   & $\pm$0.5 (cgs)  & $\pm$0.05\\
microturb. velocity & $\xi_t$     & $\pm$2\,km\,s$^{-1}$& $\pm$0.04\\
macroturb. velocity & $\xi_m$   & $\pm$2\,km\,s$^{-1}$& $\pm$0.04\\
\hline
\end{tabular}
\end{center}
\end{table}

\section{Variability analysis}\label{sect:vrbltanalysis}
\begin{figure*}
\resizebox{\hsize}{!}{\rotatebox{-90}{\includegraphics{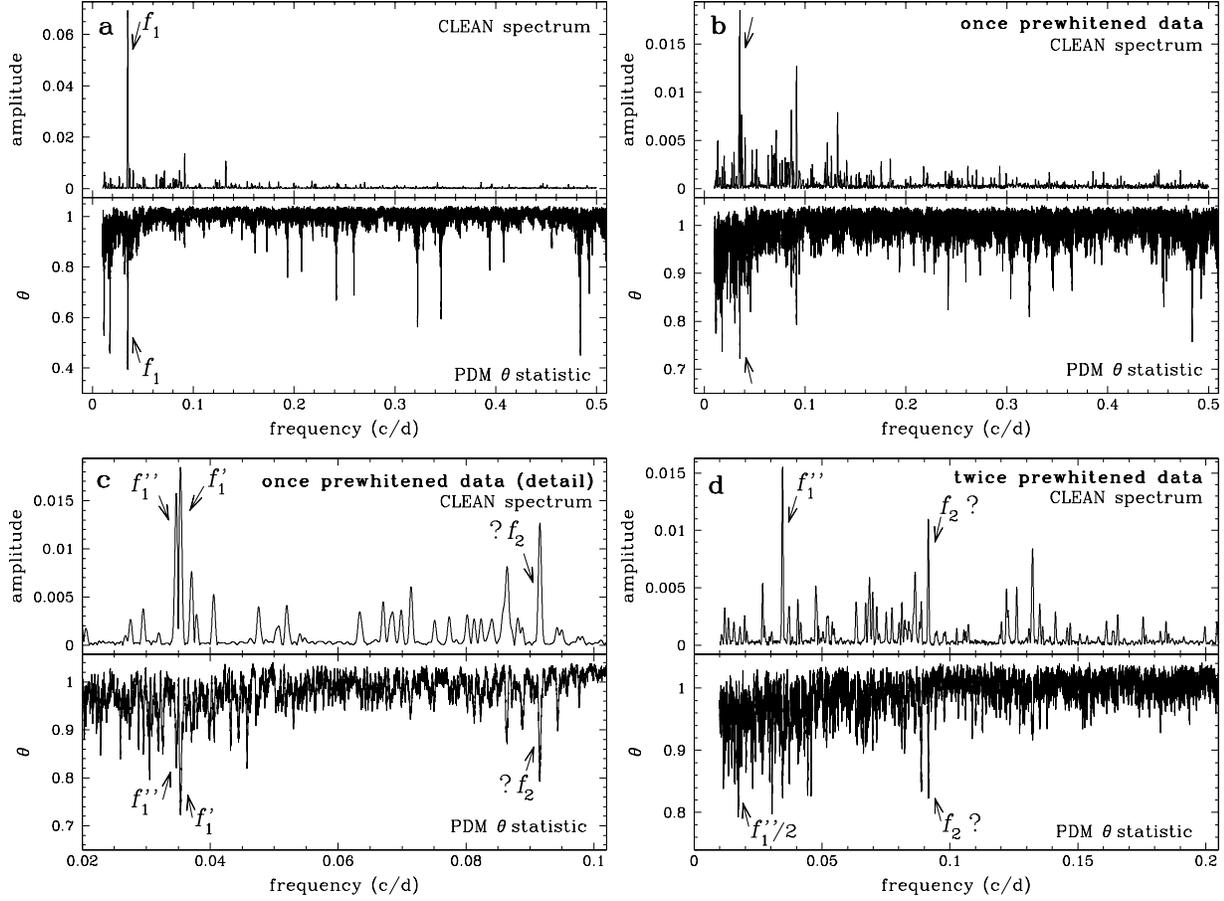}}}
\caption{The frequency analysis of the Geneva V magnitude of our 202 data
points. (a) The high resolution $\Theta$ statistic ({\sc pdm}) and the
{\sc clean}\ spectrum for the first frequency $f_1$. Both show a highly
significant peak at 0.03501\,c/d. (b) The high resolution $\Theta$ statistic
({\sc pdm}) and the {\sc clean}\ spectrum of the once prewithened data. Both
show a peak at 0.03535\,c/d. (c) Detail of the high resolution $\Theta$
statistic and the {\sc clean}\ spectrum of the once prewithened data. This
figure clearly indicates the existence of a second peak ($f''_1$), equidistant
from $f_1$ compared to $f'_1$. (d) The high resolution $\Theta$ statistic
({\sc pdm}) and the {\sc clean}\ spectrum of the twice prewithened data. The
{\sc clean} periodogram shows a peak at 0.03457\,c/d, $\Theta$ is minimal for
0.01729\,c/d, which is $\nicefrac{1}{2}$\,0.03457.}\label{fig:spectraallfr}
\end{figure*}

\object{HD\,190390} has been known to be variable for quite a long time.
According to the SIMBAD database, the variability of this star was first
mentioned by \citet{strohmeier65} who reported a photographic range of
0.4\,mag. However, dedicated observational efforts to study this variability
were made more than twenty years later. \citet{fernie86} published an analysis
based on 32 observations in the $uvby\beta(RI)$ system and mentions a range in
the $y$ magnitude of 0.3\,mag and a possible period of 28.4 or 11.8 days.
At nearly the same time, \citet{waelkens85} presented their frequency analysis
of 123 measurements in the Geneva photometric system \citep{rufener81}, obtained
with the Swiss photometric telescope at the European Southern Observatory.
The data covered 1040 days and revealed a most significant period of 28.49 days,
accounting for 80\% of the total variance. The residual scatter around
the fitted light curve was attributed to irregular cycle-to-cycle variations.

The frequency analysis was made using different algorithms: the
Jurkevich-Stelling\-werf {\sc pdm} method \citep{stellingwerf78}, the {\sc
clean} algorithm \citep{roberts87} and a general multifrequency least squares
fitting method \citep{schoenaers04}. In this analysis, we will give a rough
estimate of this error by using the formula given by \citet{montgomery99}:
\begin{equation} \label{eq:mont}
\sigma_f = \frac{\sqrt{6}\sigma_R}{\pi \sqrt{N}AT}
\end{equation}
with $\sigma_R$ the observational error, $N$ the number of measurements, $A$
the amplitude of the signal without noise, $T$ the total time span of the
observations. Different forms of this formula are given in the literature. 
\citet{schwarzenbergczerny91} introduced an extra parameter $D$ in the
formula in order to take correlations in the noise into account. To avoid an
underestimation of the calculated error, we took $T$\,=\,$N$. The motivation
is that always $N$$\le$$T$ for our different data sets of \object{HD\,190390},
because the object is measured at most only once per night. For $A$ we took in
each analysis the amplitude of the curve fitted with the found frequency. The
obtained error estimate for the proposed frequencies is only an indicative
number and should be interpreted with care. It is, however, useful to compare
frequencies from two data sets obtained with the same or a comparable
instrument. 

\begin{table}\caption{Log of the observations in the Geneva photometric system}\label{tab:genlog}
\begin{center}
\begin{tabular}{rrr|rrr}
\hline
Year &{\em N}& Time span &Year &{\em N}& Time span\\
     &   & (days) &     &   & (days) \\
\hline
1978 &  2 &  74 & 1985 & 52 & 163 \\
1979 &  1 &   - & 1986 & 24 &  77 \\
1982 & 29 &  57 & 1987 &  6 &  13 \\
1983 & 45 & 129 & 1988 & 21 & 118 \\
1984 & 22 & 128 & & & \\
\hline
\end{tabular}
\end{center}
\end{table}

\begin{table}\caption{Frequencies found in the Geneva photometry.}\label{tab:sumfr}
\begin{center}
\begin{tabular}{rrrl}
\hline
algorithm & $f$    &  P     & note\\
          & (c/d)  & (days) &     \\
\hline
\multicolumn{4}{c}{Frequency $f_1$}\\
\hline
{\sc pdm}    & 0.03501 & 28.56 & $\Theta$=0.39 \\
{\sc clean}  & 0.03503 & 28.55 & Amplitude 0.080\\
{\sc least squares fit} & 0.03502 & 28.56 & {\em Variance reduction} 62\%\\
\hline
\multicolumn{4}{c}{Frequency $f'_1$}\\
\hline
{\sc pdm}    & 0.03535 & 28.29 & $\Theta$=0.72 \\
{\sc clean}  & 0.03535 & 28.29 & Amplitude 0.033\\
{\sc least squares fit} & 0.03535 & 28.29 & {\em Variance reduction} 25\%$^*$\\
\hline
\multicolumn{4}{c}{Frequency $f''_1$}\\
\hline
{\sc pdm}    & 0.01729 & 57.84 & $\Theta$=0.79 \\
{\sc clean}  & 0.03457 & 28.93 & Amplitude 0.027\\
{\sc least squares fit} & 0.03468 & 28.84 & {\em Variance reduction} 22\%$^*$\\
\hline
\multicolumn{4}{l}{\footnotesize $^*$ on the residues}\\
\end{tabular}
\end{center}
\vskip -.5 cm
\end{table}

\subsection{Geneva photometry}\label{subsect:geneva}
The photometric data consist of 202 measurements in the Geneva photometric
system obtained with the P7 photometer on the (now decommissioned) Swiss
photometric telescope in La Silla from September 1982 to August 1988. In a
first approach we chose not to add the new Mercator measurements to this
data set for two reasons: (1) for the sake of the homogeneity of our data
set and (2) there would be a time gap of more than ten years in our data. The
Mercator measurements are discussed in a separate section
(Sect.~\ref{subsect:merc}). A small log of the observations is presented in
Table~\ref{tab:genlog}. The analysis is performed on the V magnitude. We looked
for frequencies between 0.01 and 0.5\,c/d using a frequency step of
0.00001\,c/d. 

\subsubsection{Frequency analysis}
\begin{figure}
\resizebox{\hsize}{!}{\includegraphics{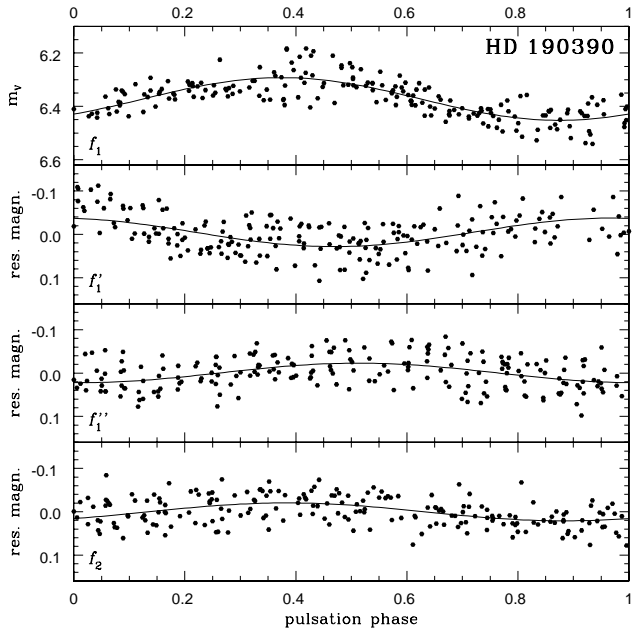}}
\caption{Phase diagrams for the visual magnitude data of \object{HD\,190390} for the
four main frequencies. Each subsequent phase diagram represents data
prewhitened with the previous frequencies.}\label{fig:fasen}
\end{figure}

All frequency analysis methods show a highly significant peak near 0.035\,c/d,
indicated by an arrow in the high resolution $\Theta$ statistic ({\sc pdm}) or
in the {\sc clean} spectrum in Fig.~\ref{fig:spectraallfr} (a). The results for
this frequency $f_1$ are summarized in Table~\ref{tab:sumfr}. In the following,
we will use for $f_1$  the value 0.03501\,($\pm$0.00001)\,c/d
(P$_1$\,=\,28.56\,d). The error estimate is obtained with formula
(\ref{eq:mont}). $f_1$ accounts for 62\% of the total variance. A phase diagram
for this frequency is shown in Fig.~\ref{fig:fasen}.

After removing (``prewhitening'') this first frequency from the data, we
performed a second frequency analysis on the residuals. We found a new
frequency very close to $f_1$: $f'_1$\,=\,0.03535\,($\pm$0.00003)\,c/d. This
frequency is documented in Table~\ref{tab:sumfr} and
Fig.~\ref{fig:spectraallfr} (b). Together with $f_1$, $f'_1$ accounts for
72\% of the total variance (an improvement of 11\%). A phase diagram for
$f'_1$ is shown in Fig.~\ref{fig:fasen}.

\begin{figure}
\resizebox{\hsize}{!}{\rotatebox{-90}{\includegraphics{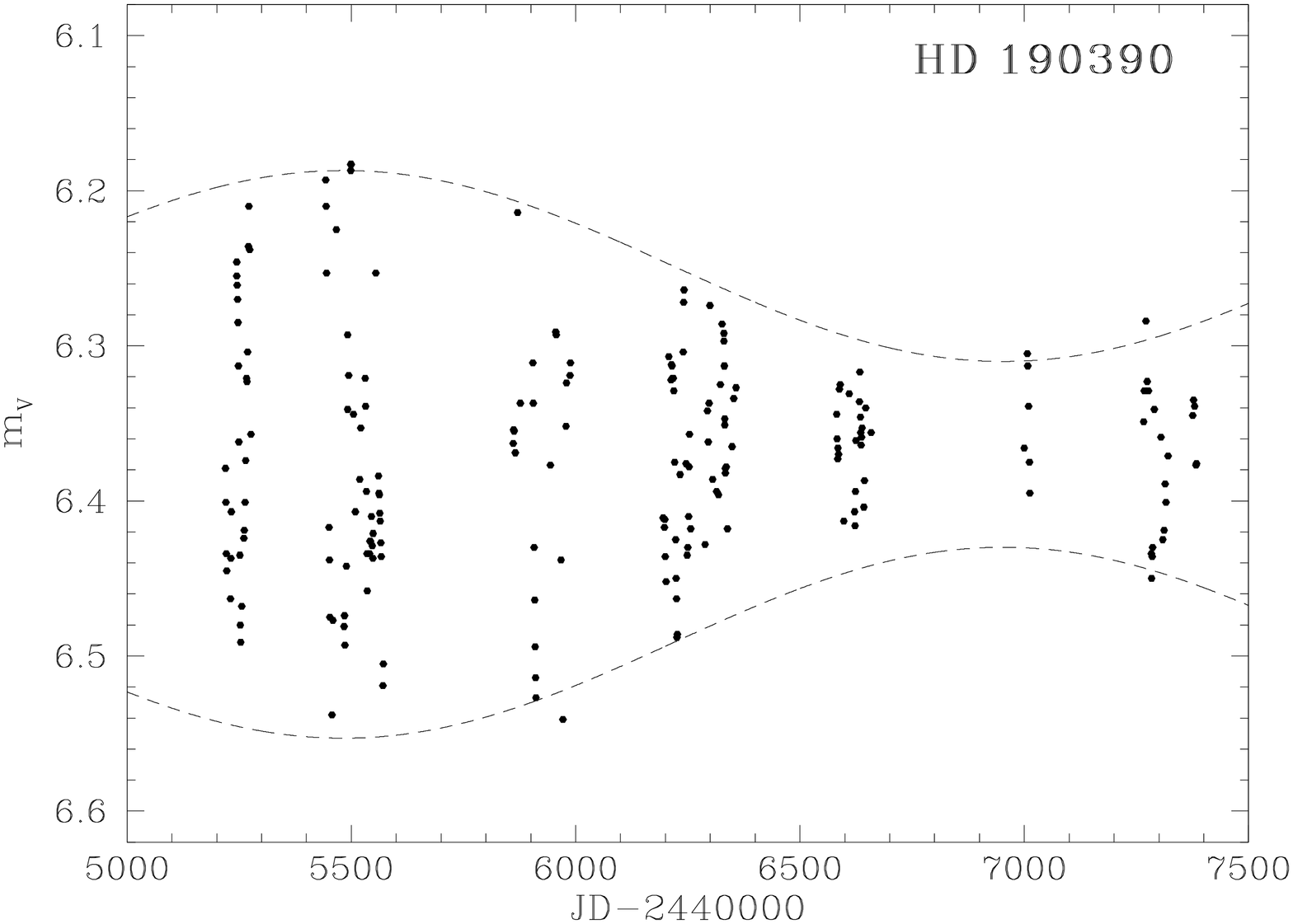}}}
\caption{The Geneva V magnitude together with a harmonic fit with frequency
$f_b$\,=\,$f'_1$$-$$f_1$\,=\,0.00034\,c/d through the extrema (either minimum
or maximum) of each observational season.}\label{fig:genevadat}
\end{figure}

In the periodograms of the once prewhitened data (Fig.~\ref{fig:spectraallfr},
b), we note, apart from the peak at 0.03535\,c/d ($f'_1$), another peak very
close to the main frequency $f_1$, situated at 0.03469\,c/d. This second peak
can easily be seen on Fig.~\ref{fig:spectraallfr} (c), which is a zoom-in of
panel (b). The peak at 0.03469\,c/d is at the position $f_1-(f'_1-f_1)$ and
therefore it could be the third component of a frequency {\em triplet},
together with $f_1$ and $f'_1$. A natural question is then if this frequency is
also visible in an analysis of the twice prewhitened data. The answer is
positive for the {\sc clean} spectrum, which has a maximum at
$f''_1$\,=\,0.03457\,($\pm$0.00005)\,c/d. $\Theta$ is minimal for
$f$\,=\,0.01729\,c/d, which is $\nicefrac{1}{2}$$f''_1$ (see
Fig.~\ref{fig:spectraallfr}, d). $f''_1$ accounts for an additional 6\% of
the variance (from 72\% to 78\%). Other details can be found in
Table~\ref{tab:sumfr}. A phase diagram for this frequency is shown in
Fig.~\ref{fig:fasen}.

A strong argument for the presence of additional frequencies is given by a
simple inspection of Fig.~\ref{fig:genevadat}. If $f'_1$ and/or $f''_1$ are
physical, then a {\em beating} should be observed with frequency
$f_b$\,=\,$f'_1$$-$$f_1$\,=\,0.00034\,c/d or with a period of
P$_b$\,$\simeq$\,3000\,d. This beating can be seen in Fig.~\ref{fig:genevadat}:
the peak-to-peak amplitude {\em per year} clearly varies, reaching its maximum
in 1983 (0.355\,mag) and its minimum in 1986 (0.099\,mag). The dashed line is a
simple harmonic fit with frequency $f_b$ through the extrema (either minimum or
maximum) of each year in which observations have taken place.

In the periodograms of the once prewhitened data, a prominent peak is also seen
at 0.09154\,c/d (Fig.~\ref{fig:spectraallfr}, b and c). The same peak is also
present in the twice prewhitened data (Fig.~\ref{fig:spectraallfr}, d). Since
this frequency is not linked to the frequencies around 0.035\,c/d, we can
consider this frequency as a genuine second frequency
$f_2$\,=\,0.09154\,($\pm$0.00005)\,c/d. $f_2$ is also the highest peak in the
{\sc clean} periodogram after removing the three frequencies around 0.035\,c/d. 
A phase diagram for this second frequency is shown in Fig.~\ref{fig:fasen}.

\subsubsection{Light curves and colour variations}
\begin{figure}
\resizebox{\hsize}{!}{\includegraphics{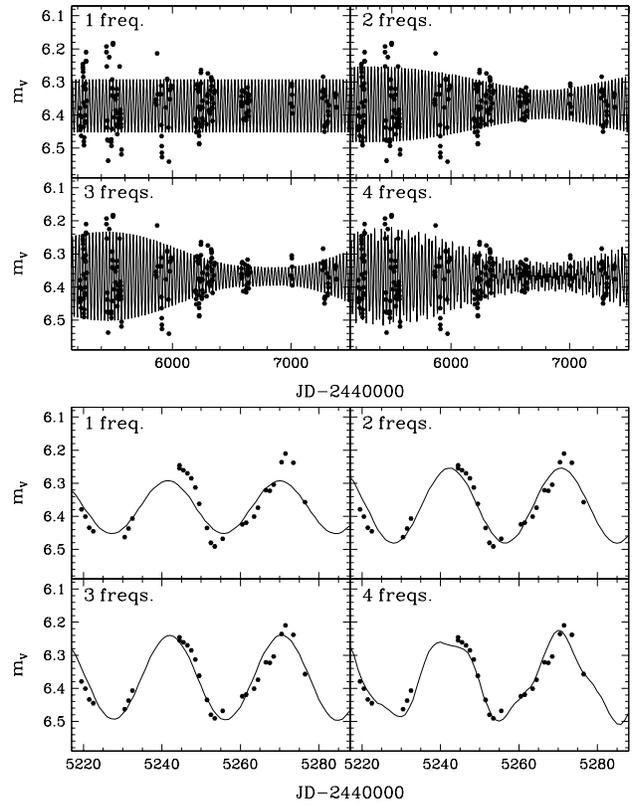}}
\caption{Light curve of \object{HD\,190390}. The light curve is constructed
using the number of frequencies indicated in the upper left corner of each
panel on the figure. {\em upper panel}: global behaviour of the light curve. It
is clear that the beating which was already noted in Fig.~\ref{fig:genevadat}
is sampled quite well when making use of more than one frequency. On the other
hand, the overall amplitude of the beating seems to be underestimated.
{\em lower panel} Local behaviour of the light curve of \object{HD\,190390}.}\label{fig:f1h3f2f3}
\end{figure}

A light curve was constructed using the four frequencies we found above. The
fits were obtained using a simple least-squares procedure and the results can
be seen in Fig.~\ref{fig:f1h3f2f3}. In the upper panel of this figure, we show
how the light curve behaves {\em globally}. The beating is present in the light
curve, even when making use of only two frequencies. However, the
{\em amplitude} of the beating seems to be underestimated. In this respect, the
inclusion of the fourth frequency ($f_2$) improves the global behaviour of the
light curve. In the lower panel of Fig.~\ref{fig:f1h3f2f3}, a detail of the
light curve between JD\,2445217 and JD\,2445288 is shown.

\begin{figure}
\resizebox{\hsize}{!}{\includegraphics{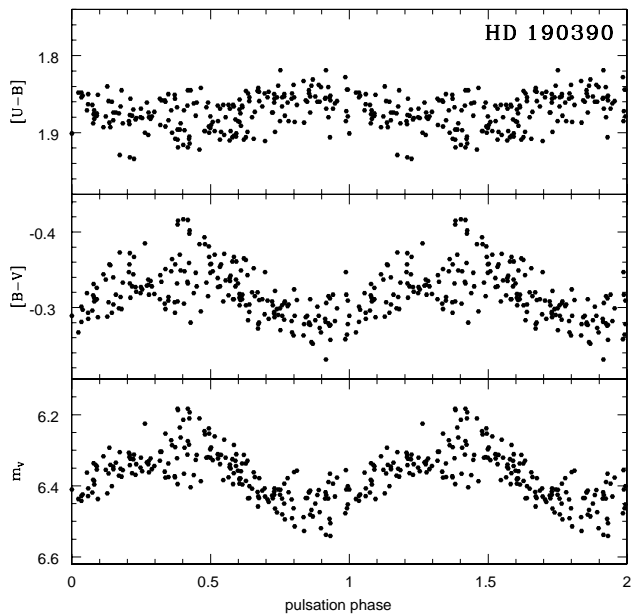}}
\caption{The U-B and B-V colour variations folded on the same period (28.56\,d)
as the light variations (m$_{\rm V}$). The displayed colour variations are
clearly in antiphase respectively phase with the light variations.}
\label{fig:lghtclrsfolded}
\end{figure}

The frequency analysis presented above was based on the V magnitude because
this magnitude is mostly used in this kind of analysis. In
Fig.~\ref{fig:lghtclrsfolded} the colour variations and the light variations
are folded on the same period. As was noted by \citet{waelkens85}, the U-B
and B-V variations are respectively in antiphase and in phase with the light
variations. This result is important in the discussion of the pulsational nature
of \object{HD\,190390} (Sect.~\ref{subsect:uhrclassification} and
\ref{subsect:wvrclassification}).

\subsection{HIPPARCOS data}\label{subsect:hipp}
53 HIPPARCOS measurements of \object{HD\,190390} were extracted from the
HIPPARCOS catalogue \citep{ESA97}. 43 measurements had a non-zero quality flag,
but at 10 more epochs the star was 0.10 to 0.15 mag brighter than on average.
We have no explanation for this brightening, but since we cannot exclude that
these brightenings are instrumental, we used only the remaining 34 points. They
cover a total time span of 948.6 days, from April 1990 to November 1992. The
precision of these HIPPARCOS magnitudes (which is also given in the catalogue
for each measurement individually) is between 0.005 and 0.01\,mag.

It is remarkable that also in this very limited data set, the second best
two-frequency solution is the combination of $f_{\rm H1}$\,=\,0.03704 c/d
(P$_{\rm H1}$\,=\,27.00\,d) and $f_{\rm H2}$\,=\,0.09042\,c/d (P$_{\rm
H2}$\,=\,11.06\,d). These frequencies account for 95.8\% of the total variance.
The best combination in the least-squares sense, $f_{\rm H1'}$\,=\,0.07738 c/d
(P$_{\rm H1'}$\,=\,12.92\,d) and $f_{\rm H 2'}$\,=\,0.09044\,c/d (P$_{\rm
H2'}$\,=\,11.06\,d) does not give a significantly different result
(96.4\% variance reduction). The frequencies  $f_{\rm H1}$ and $f_{\rm H1'}$
could be related aliases since the second one is not recovered after
prewhitening with the first frequency and vice-versa, and there is no doubt
that the star has a main period of around 28 days. 

\subsection{CORALIE radial velocities}\label{subsect:snel}
The CORALIE fiber-fed echelle spectrograph at the Swiss 1.2\,m telescope at
La Silla has a resolution of $\sim$50000 at 5000\,\AA\ and a wavelength
coverage from 3880 to 6810\,\AA. Due to the stability of the whole
configuration and the possibility of making simultaneously Thorium-Argon
calibration spectra in the second fibre, the spectrograph is specifically
designed for high-precision radial-velocity measurements.

\begin{figure}
\resizebox{\hsize}{!}{\includegraphics{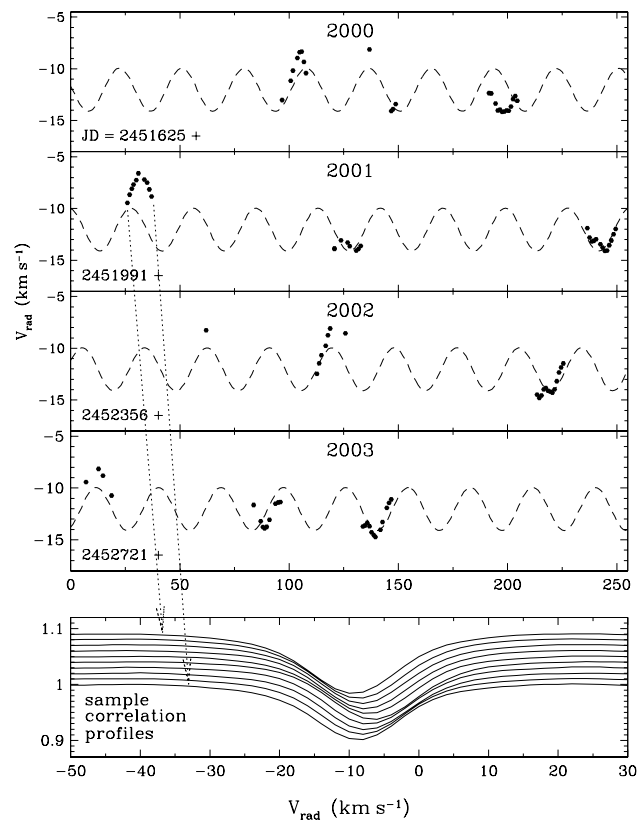}}
\caption{The CORALIE radial velocity data from 2000 to 2003. Several groups
of points can be discerned, indicating the different observational runs.
A radial velocity curve with a period of P\,=\,28.56\,days is fitted to the
data. This period was found in the photometric data, and also in the
{\sc clean} analysis of the radial velocity data. At the bottom, sample
correlation profiles are given for one observational run.}\label{fig:smcorwim}
\end{figure}

\begin{figure}
\resizebox{\hsize}{!}{\includegraphics{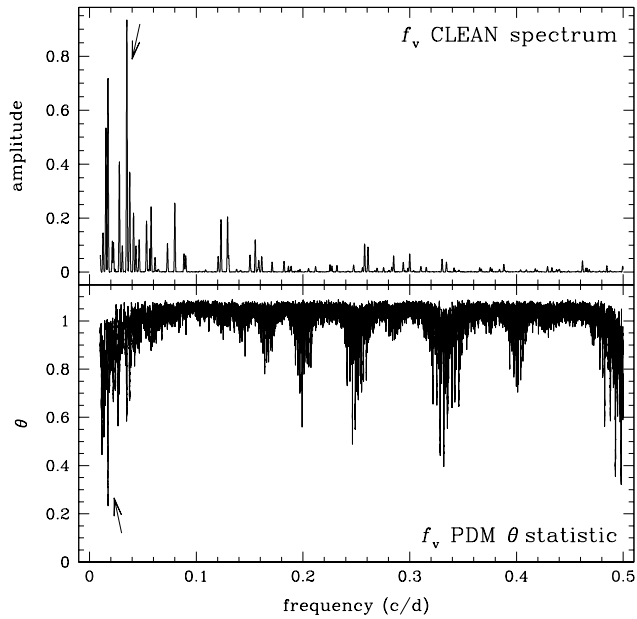}}
\caption{The high resolution $\Theta$ statistic ({\sc pdm}) and the {\sc clean}
spectrum of the CORALIE radial velocities. {\sc clean} shows a peak at
0.03502\,c/d, while the {\sc pdm} $\Theta$ statistic is minimal for
$f$\,=\,0.01745\,c/d.}\label{fig:snelspectra1stefr}
\end{figure}

Since November 1999, 103 radial velocity measurements of \object{HD\,190390}
have been collected (Fig.~\ref{fig:smcorwim}). Velocities are deduced by
cross-correlating the spectra with an appropriate mask. In the case of
\object{HD\,190390}, an F0 mask is used. Sample correlation profiles for one
observational run are given at the bottom of Fig.~\ref{fig:smcorwim}. The
accuracy of these velocities is $\sim$1.0--1.5\,km\,s$^{-1}$. From
Fig.~\ref{fig:smcorwim} it is clear that the period in the radial velocity
data must be close to the main period $f_1$ found in the photometric data. The
frequency found in the frequency analysis (Table~\ref{tab:sumsnel1stefr} and
Fig.~\ref{fig:snelspectra1stefr}) is indeed equal to the main frequency found
in the Geneva photometry.

\begin{table}\caption{Frequency analysis of the CORALIE radial velocity data.
Note that both {\sc pdm} and the {\sc least squares} method find
$\nicefrac{1}{2}f_1$ instead of $f_1$.}\label{tab:sumsnel1stefr}
\begin{center}
\begin{tabular}{rrrl}
\hline
algorithm & $f_{\rm v}$  &  P$_{\rm v}$ & note\\
          & (c/d)  & (days) &     \\
\hline
{\sc pdm}    & 0.01745 & 57.31 & $\Theta$=0.23 \\
{\sc clean}  & 0.03502 & 28.56 & Amplitude 0.93 \\
{\sc least squares fit}& 0.01741 & 57.44 & {\em Variance reduction} 76\% \\
\hline
\end{tabular}
\end{center}
\end{table}

\begin{figure}
\resizebox{\hsize}{!}{\includegraphics{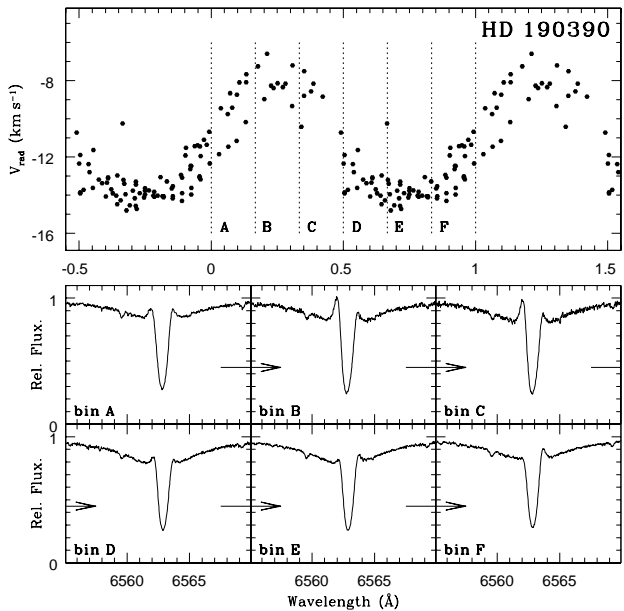}}
\caption{{\em upper panel} Phase bins used in the co-adding of the 103 CORALIE
spectra. In this ``optimized'' phase diagram with frequency 0.03501\,c/d, a
specific phase was determined for each observing run separately. {\em lower
panel} The H$\alpha$ line in the six resulting spectra showing a different
profile during one cycle.}\label{fig:phbns}
\end{figure}

Variation of the H$\alpha$ profile during a pulsational cycle is often
used as a powerful diagnostic tool in the determination of the pulsation
character. The CORALIE spectra, however, are of too low signal-to-noise (s/n)
for a detailed profile variation study, because the first intention of these
spectra was only to obtain a radial velocity by cross-correlation. For this
purpose, a s/n of around 30 is sufficient, and the mean s/n of all spectra is
indeed 40. Nontheless, we {\em can} study line profile variations over a cycle
by combining spectra in the same phase bin. In such a procedure, a limited
number of higher s/n spectra is compiled out of spectra situated in the same
phase bin. In the construction of these combined spectra, we chose to use 6
phase bins designated A to F, illustrated in Fig.~\ref{fig:phbns}. The number
of spectra added, the total integration time and the achieved s/n ratios are
summarized in Table~\ref{tab:phsbns}. Spectra were first transformed to rest
wavelength, normalised and then added, s/n weighted.

From Fig.~\ref{fig:phbns} it is clear that H$\alpha$ in \object{HD\,190390} is
variable during a pulsational cycle: relatively strong emission in the blue
wing is seen in bins B and C, while in bin F for example, there is only some
emission present in the red wing, and none in the blue. H$\alpha$ emission in
the blue wing was already noted by \citet{sasselov85}, and later confirmed by
\citet{luck90}. \citeauthor{sasselov85} gave a $\sim$60\,km\,s$^{-1}$
separation between the emission and the photospheric rest velocity of the
H$\alpha$ absorption. We find similar values for this separation, but the exact
value for each phase binned spectrum depends on the method adopted to
determine the central wavelength of both the emission and the absorption peak.

\begin{table}\caption{Details of the phase binning of the 102 CORALIE spectra
into 6 bins. The older measurement of 1999 was not used.}\label{tab:phsbns}
\begin{center}
\begin{tabular}{rrrrr}
\hline
\multicolumn{2}{r}{Phase bin} & \# spectra & total integra- & achieved s/n \\
   &                          &            & tion time (s)  & (at 6550\,\AA)\\
\hline
 A &$-$0.083 - 0.083 & 14 & 2968 & 160 \\
 B &   0.083 - 0.250 & 10 & 1713 & 100 \\
 C &   0.250 - 0.417 &  6 & 1200 &  90 \\
 D &   0.417 - 0.583 & 22 & 5632 & 150 \\
 E &   0.583 - 0.750 & 25 & 7343 & 180 \\
 F &   0.750 - 0.917 & 25 & 7172 & 150 \\
\hline
\end{tabular}
\end{center}
\end{table}

\subsection{Geneva photometry with Mercator}\label{subsect:merc}
The Mercator telescope is a 1.2 meter telescope located on the Roque de los
Muchachos observatory on La Palma, Spain. The Mercator telescope, operated by
our Institute, is a twin of the Swiss telescope in La Silla. Two instruments
are available for the moment: the refurbished photometer P7 which was in use at
the ``old'' Swiss telescope at La Silla, and a CCD camera.

\object{HD\,190390} is in the standard program catalogue for P7, but since the
declination of this object is rather low (-11$^{\circ}$36$^m$), the minimum
air mass for La Palma is quite high ($F_z$\,=\,1.30). Therefore,
\object{HD\,190390} is always measured in concatenation with the standard star
\object{HD\,190172} (F4III, V=6.72) and the comparison star \object{HD\,192771}
(G5, V=8.7). 72 measurements of \object{HD\,190390} have been collected with the
Mercator telescope. During the reduction, a quality flag is assigned to each
observation, ranging from 0 to 4. Observations with the lowest quality (flag 0)
were discarded, and 66 measurements remained, covering 787 days. This data set
is still quite limited to perform a reliable frequency analysis. However, a
preliminary analysis again suggests a frequency close to the main frequency
found in the ``old'' Geneva data:
$f_{\rm M}$\,=\,0.03524\,c/d\,($\pm$0.00005)\,c/d
(P$_{\rm M}$\,=\,28.38\,days). This frequency accounts for 86\% of the total
variance.

There is a secondary peak in the periodograms near the
main frequency and at 0.0897\,c/d an unrelated frequency is found. The best
3-frequency solution (94\% reduction of the initial variance) is with 0.03521,
0.04072 and 0.08970\,c/d.

\begin{figure}
\resizebox{\hsize}{!}{\rotatebox{-90}{\includegraphics{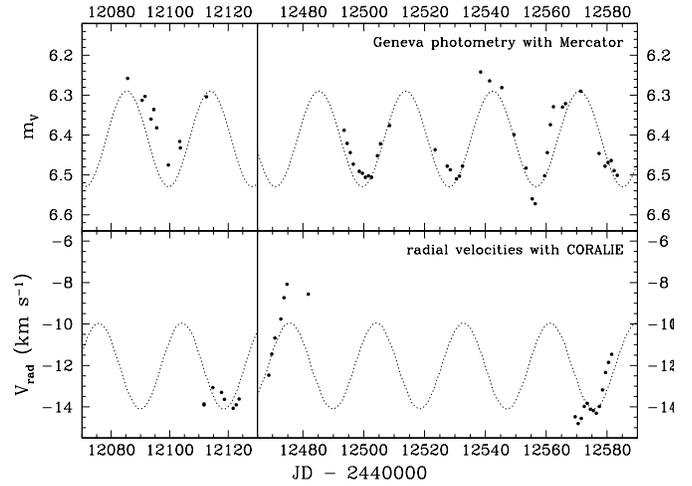}}}
\caption{{\em upper panel} Overview of the Geneva photometry collected with the
Mercator telescope, fitted with a harmonic with the main frequency
(0.03501\,c/d). In the {\em lower panel} a comparison is made with the CORALIE
radial velocities of the same period. The dotted line is again a harmonic with
the main frequency. It is clear that both curves are mirror images from each
other.}\label{fig:merc}
\end{figure}

\subsection{Variability}\label{subsect:puttogether} 
\subsubsection{Photometry and velocity}
In Fig.~\ref{fig:merc} we compare the light curve of \object{HD\,190390}
with its radial velocity curve. The comparison is made for a time interval
for which we have both photometric and velocity data, so no extrapolation is
needed. Both curves are fitted with a simple harmonic with the main frequency
(0.03501\,c/d). It is clear that both curves are mirror images from each other.

\subsubsection{Beating in the Mercator data}
\begin{figure}
\resizebox{\hsize}{!}{\includegraphics{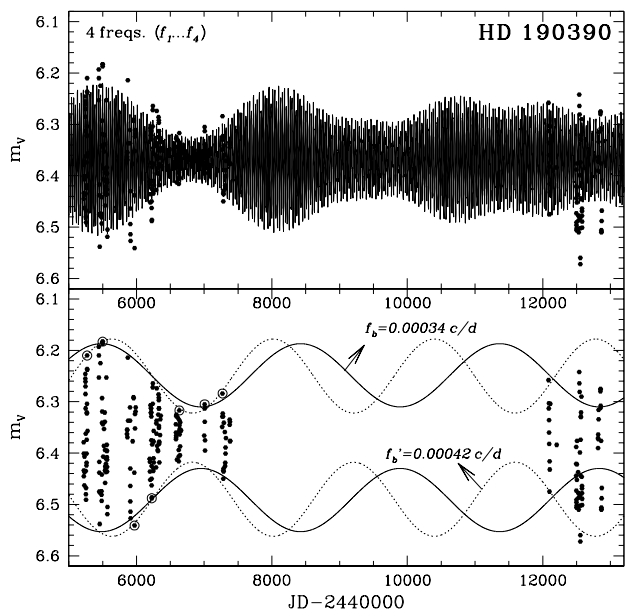}}
\caption{Comparison of the new Mercator photometry and the ``old'' Geneva
photometry. In the {\em upper panel} the light curve of \object{HD\,190390}
constructed with four frequencies is extrapolated to the time range covered by
the Mercator data (points right on the figure). Extrapolation of the light
curve is, however, far from robust, since a small change in frequency will
cause a rather large change in the global behaviour of the light curve in the
extrapolated region. In the {\em lower panel} a harmonic fit with frequency
$f_b$\,=\,$f_2$$-$$f_1$\,=\,0.00034\,c/d (full line) and frequency 
$f'_b$\,=\,0.00042\,c/d (dotted line) through the extrema (marked on the figure
with an open dot) of each year (1982--1988) is constructed. The latter frequency
is the best fit {\em through the extrema}, while the former was found in our
frequency analysis (see Table~\ref{tab:sumfr}).}\label{fig:beatmerc}
\vskip -.5 cm
\end{figure}

To determine whether the beating is seen in the Mercator data, we extended the
light curve constructed in Fig.~\ref{fig:f1h3f2f3} to the time range covered by
the Mercator data (upper panel of Fig.~\ref{fig:beatmerc}). We remind that the
amplitude of the light curve seems to be underestimated. Thus, the points of
the second observational run on Mercator being quite off the light curve (by
$\sim$0.12\,mag) is not an argument to reject a stable beating. Caution is
needed because a small difference in the frequencies will cause a quite large
difference in the {\em global} behaviour of the extrapolation.

Another and more robust method is depicted in the lower panel of the same
figure. There, a simple harmonic fit was constructed through the extreme
values (either minimum or maximum) of each year (1982--1988) and with a
fixed frequency of $f_b$\,=\,$f'_1$$-$$f_1$\,=\,0.00034\,c/d. This fit is
then mirrored with respect to the mean m$_{\rm V}$ value, and an enveloping
curve is obtained. The data points of the second Mercator run again fall out
of the permitted region, so they do not seem to be compatible with the beating.
However, with a small change in the beating frequency
($f_b'$\,=\,0.00042\,c/d), we were able to make these points ``permitted''.

Thus, a beating phenomenon is not excluded although the values of the
frequencies around the main frequency seem not well defined. If a very close
duplet or triplet of frequencies is present, the total time span of the
observations could still be insufficient to resolve the exact frequency values.
Alternatively, it is not excluded that the main frequency $f_1$ changes its
value {\em and} its amplitude in time in a more or less random way. In this
case the nearby frequencies are only ``remnants'' of the broadened peak in the
periodogram not dealt with by the prewhitening.

For the secondary frequency, near 0.09\,c/d, we found indications of its
presence in independent data sets, and it might correspond to the
frequency $f$\,=\,0.0847\,c/d that was already found by \citet{fernie86}. Here
also its value is either variable in time or some beating of closely spaced,
so-far unresolved frequencies is present.

\section{Discussion}\label{sect:dscssn}

\subsection{Luminosity}\label{sect:lmnst}
In this first section of the discussion, the luminosity of \object{HD\,190390}
is estimated. As a first step, the reddening of the object is quantified by
calculating synthetic colour indices. Then, two different methods are used to
obtain the luminosity of our object.

\noindent{\bf Synthetic colour indices and reddening}
The reddening is estimated by comparing the observed Geneva colours of our
object with calculated synthetic colours. Such synthetic colours were calculated
by combining the digitized passband of the Geneva filters \citep{rufener88}
with the Kurucz model atmospheres \citep[see][for more details]{reyniers02a}.
The synthetic B-V of \object{HD\,190390} with model (T$_{\rm eff}$, $\log g$,
[M/H]) = (6250\,K, 1.0, $-$1.5) is B-V\,=\,$-$0.574. The mean B-V in our Geneva
data is B-V\,=\,$-$0.318, hence E(B-V)\,=\,0.256\,$\pm$\,0.05\,mag. This value
is somewhat higher than the value derived by \citet{fernie86}:
E(B-V)\,=\,0.10\,mag.\\
{\bf Luminosity with trigonometric parallax}
The parallax found in the HIPPARCOS catalogue (retrieved via SIMBAD) is
$\pi$\,=\,2.56\,($\pm$0.97)\,mas, which gives a distance of d\,$=$\,391\,pc,
with an upper limit of 629\,pc and a lower limit of 283\,pc.
Taking the reddening into account, we get an absolute visual magnitude of
M$_{\rm V}$\,$=$\,$-$2.29. However, the uncertainty induced by the parallax is
much larger than the uncertainty of the reddening:\\
\phantom{aa} d\,$=$\,629\,pc $\rightarrow$ M$_{\rm V}$\,$=$\,$-$3.33\\
\phantom{aa} d\,$=$\,283\,pc $\rightarrow$ M$_{\rm V}$\,$=$\,$-$1.60\\
Assuming the bolometric correction to be negligible
\citep[see e.g.][]{flower96}, we derive a luminosity of
L\,$=$\,656\,L$_{\odot}$. An upper limit is obtained for this method by taking
d\,$=$\,629\,pc, which yields L\,$=$\,1701\,L$_{\odot}$ or
log(L/L$_{\odot}$)\,$=$\,3.2.\\
{\bf Luminosity with P-L-[Fe/H] relation}
\citet{nemec94} derived P-L-[Fe/H] relations and distances for several classes
of Pop.~II variables. Given the pulsational properties of \object{HD\,190390}
together with its metal deficiency, there is strong evidence that this object
belongs to the class of Pop.~II Cepheids (see also
Sect.~\ref{subsect:wvrclassification}).
The P-L-[Fe/H] relation for this class depends on the pulsation mode (either
fundamental or first-overtone). With the P-L-[Fe/H] relations given in
\citet{nemec94}, and P\,$=$\,28.56\,d, we obtain\\
\phantom{aa} M$_{\rm V}$(F)\,$=$\,$-$2.47\,($\pm$\,0.07) \phantom{a} {\em (fundamental)}\\
\phantom{aa} M$_{\rm V}$(H)\,$=$\,$-$2.92\,($\pm$\,0.07) \phantom{a} {\em (first overtone)}\\
where the error of 0.07 only accounts for the error on the P-L-[Fe/H] relation.
This error is, again, probably underestimated, not only because the variables
in the formula are uncertain (especially [Fe/H]), but also because the
relation was derived with data from cluster variables. It is therefore not
clear if this relation also holds for field stars like \object{HD\,190390}.
Note also that with a maximum magnitude of M$_{\rm V}$\,$=$\,$-$3, we derive a
distance of d\,$=$\,673\,pc (log(L/L$_{\odot}$)\,$=$\,3.1 and distance from the
galactic plane $|Z|$\,=\,247\,pc).

\subsection{UU~Her classification}\label{subsect:uhrclassification}
UU~Her type stars are defined by \citet{sasselov84} as variable F-type
supergiants at high galactic latitudes having long periods and low amplitudes.
Further characteristics of this group are \citep{bartkevicius92}:\\[-.65 cm]
\begin{itemize}
\item metal deficiency
\item high velocities
\item very specific variability with relatively small amplitudes and two or
three alternating periods form 20 to 120 days
\item large infrared (IR) excesses due to circumstellar dust
\end{itemize}
\vskip -.2 cm
However, an object does not have to possess all properties mentioned here
to be a member of this heterogeneous class. The evolutionary status of these
stars is not yet clear, but most characteristics clearly point to a final stage
of high luminosity or a post-AGB stage.

Given its low metal content already mentioned by \citet{mcdonald76} and its
high galactic latitude, \object{HD\,190390} was mentioned as a possible UU~Her
candidate \citep{waelkens85, sasselov85, fernie86}. However, due to the
apparently stable and short period (relative to other UU~Her candidates) found
by \citeauthor{waelkens85}, the genuine character of \object{HD\,190390} as
UU~Her star was doubted. Moreover, \object{HD\,190390} has a ``normal'' radial
velocity and no significant infrared excess has been observed by IRAS.

With the observational material presented here, we are able to further contest
this UU~Her classification. We confirm the stable \citeauthor{waelkens85}
period of $\sim$28.6\,d. Our radial velocity data further confirm the
``normal'' velocity of \object{HD\,190390}. Also the fact that there is no
significant reddening is opposite to what is seen in most UU~Her type stars.
Therefore, a {\em post-AGB stage} of our object is less likely. Also,\\[-.65 cm]
\begin{itemize}
\item the chemical signature does {\em not} point to a 3rd dredge-up scenario
(no carbon or s-process enhancement)
\item the (however very uncertain) luminosity (not higher than
log\,$L/L_{\odot}$\,$\simeq$\,3.2) is too low to support a post-AGB stage
\end{itemize}
\vskip -.2 cm
The absence of a clear 3rd dredge-up signature is, however, not enough
to exclude a post-AGB scenario \citep[e.g.][]{vanwinckel97}.

\subsection{W~Vir classification}\label{subsect:wvrclassification}
W~Virginis variables are pulsating stars of population II located in the Cepheid
instability strip. In the HR diagram, W~Vir stars occupy a region slightly
lower and slightly to the right compared to the classical (Pop.~I) Cepheids.
Periods are between 12 and about 20 days and amplitudes between 0.3 and
1.2 mag. W~Vir stars are in fact part of the larger class of Pop.~II Cepheids,
containing also the BL Her stars at the shorter period end (P\,=\,1$-$5\,d), and
the RV Tauri stars at the longer period end (P\,=\,20$-$100\,d); see
\citet{wallerstein02} for a review of Pop.~II Cepheids. Spectra of W~Vir stars
sometimes show hydrogen and helium emission near maximum light, and
double metallic absorption lines, usually explained by the propagation of a
shock wave through the photosphere.

The metal deficiency, the luminosity (too low for post-AGB) and most importantly
the length of the period itself suggest a W~Vir classification for
\object{HD\,190390}. Furthermore, the CNO chemical pattern points to a 1st
(post-RGB) rather than a 3rd (post-AGB) dredge up. The (anti-)correlations seen
in the colours are reminiscent of variables in the instability strip. Moreover,
the luminosity inherent to this classification is compatible (within the error)
with the luminosity derived by the trigonometric parallax. On the other hand,
the spectroscopic features seen in \object{W~Vir} (strong hydrogen and helium
emission and double metallic absorption lines) are not seen in the spectrum
(except the weak emission in the wing of H$\alpha$). The photosphere of
\object{HD\,190390} is seemingly much more stable than that of \object{W~Vir},
and no strong shock wave is developed during the pulsation.

\subsection{Lithium in \object{HD\,190390}}
In Sect.~\ref{sect:abndc} we derived a Li abundance of
log\,$\epsilon$(Li)\,=\,1.9. For an extensive discussion of Li in evolved
stars, we refer to Sect.~5.2 in \citet{reyniers01}. The Li abundance of
\object{HD\,190390} is {\em not} higher than the interstellar Li abundance of
log\,$\epsilon$(Li)\,=\,3.1. However, since the initial Li abundance of
\object{HD\,190390} (thought to be the Spite plateau) is approximately the
present one, and since at least one mixing event has occurred in
\object{HD\,190390} (as seen from the CNO pattern), the Li content is expected
to be much lower than the initial one. Note that a Li dilution of a factor of
20 during the 1st dredge-up is expected for stars in this metallicity range
\citep[e.g.][]{gratton00}. Therefore, Li {\em production} is required in the
case of \object{HD\,190390}.

An obvious Li producing scenario for \object{HD\,190390}, which was also
invoked by \citet{luck90}, is a ``hot bottom burning'' (HBB) scenario in which
the base of the convective envelope of a AGB star is hot enough for the
hydrogen burning to ignite and in which fresh $^7$Li is quickly transported to
cooler layers before it is destroyed by proton capture. However, only AGB stars
of intermediate mass (4-5\,M$_{\odot}$) or equivalently high luminosity
M$_{\rm bol}$\,$\la$\,$-$6 are thought to exhibit this HBB, and therefore such
a scenario is difficult to reconcile with a W~Vir or post-AGB classification.
There exists also a class of Li enriched galactic low mass field Carbon stars
\citep{abia93, abia97, abia00}, but these stars are chemically very different
from \object{HD\,190390}.

Turning to less evolved stars, Li enrichment is also seen in about one percent
of ``normal'' G-K giants. Some of these giants show Li abundances even larger
than that of the interstellar medium. While the exact mechanism responsible
for this enrichment is still not known, internal production of $^7$Li is also
for this group of stars most favourable \citep[e.g.][]{delareza97}. Non-standard
mixing models are developed to explain these high Li abundances and commonly
dubbed ``cool bottom processing'' \citep[CBP,][]{wasserburg95,sackmann99}. An
attractive but highly speculative scenario for the Li enhancement in
\object{HD\,190390} could be a scenario in which \object{HD\,190390} turned
into a Li-rich K giant during its ascent on the red giant branch and in which
it was able to preserve this Li in its outer layers.

The 25.7\,day Population II Cepheid V42 in the globular cluster M5
\citep{carney98} is a metal poor ([Fe/H]\,=\,$-$1.22) star with a surprisingly
high Li abundance of log\,$\epsilon$(Li)\,$\simeq$\,1.8. No intrinsic
neutron-capture element enhancement was observed either. Using the distance to
M5, the authors give an estimate for the mean absolute visual magnitude of
M$_V$\,=\,$-$3.15. Apart from these characteristics very similar to
\object{HD\,190390}, some differences are also seen. Oxygen for example is
slightly deficient in this star, whereas sodium is enhanced. This could be
important evidence that deep mixing has taken place, in which products of the
ON part of the CNO cycle are brought from the hydrogen-burning shell to the
surface.

\section{Conclusion}\label{sect:cnclsns}
In this paper an attempt was made to clarify the evolutionary status of the
luminous F-type supergiant \object{HD\,190390}. The star was previously
classified as a post-AGB star, based on its high galactic latitude and metal
deficiency. However, no unambiguous {\em chemical} indications were found
(carbon and s-process enrichment) to support this view.

Here, we confirm the main pulsational period of $\sim$28.6\,d. The high
resolution, high signal-to-noise spectra confirmed the metal deficiency, the
C deficiency, the N and O enhancement and the unexpectedly high Li abundance.
A detailed abundance study of the s-process elements indicated a slight
s-process enhancement which is in our view probably not intrinsic, but a
consequence of the (small) scatter in the galactic chemical evolution. We were
not able to give a conclusive evolutionary status for this star, but contrary
to the post-AGB classification found in the literature, we suggest a W~Vir
classification instead.

Analysis of the different independent data sets showed that in this star a
second periodicity is present near 11.1\,d. We suggest that the pulsation
has a beating with frequency $f_b$\,=\,0.00034\,c/d. This could be a working
hypothesis guiding future observations of \object{HD\,190390}.

\begin{acknowledgements}
MR acknowledge financial support from the Fund for Scientific Research -
Flanders (Belgium). This research has made use of the Vienna Atomic Line
Database (VALD), operated at Vienna, Austria, and the SIMBAD database,
operated at CDS, Strasbourg, France. Hans Van Winckel, Peter De Cat and Katrien
Kolenberg are warmly thanked for many improvements on early versions of the 
manuscript.
\end{acknowledgements}

\bibliographystyle{aa}

\end{document}